\documentclass[12pt,onecolumn]{IEEEtran}
	\IEEEoverridecommandlockouts
	\usepackage{amsfonts}
	\usepackage{amssymb}
	\usepackage{amscd}
	\usepackage{graphics}
	\usepackage[dvips]{graphicx}
	\usepackage{epsfig}
	\usepackage{subfigure}
	\usepackage{amsmath}
	\usepackage{array}
	\usepackage{eqparbox}
	\usepackage[bookmarks=false]{hyperref}
	\usepackage{fancyhdr}
	\usepackage{mathrsfs}
	\usepackage[shortlabels]{enumitem}
	\renewcommand{\baselinestretch}{1.58}

	\newtheorem{thrm}{Theorem}
	\newtheorem{prop}{Proposition}
	
	\newtheorem{defn}{Definition}

\begin{document}
	\title{On The Equivalence of Projections In Relative $\alpha$-Entropy and R\'{e}nyi Divergence}
	\author{\IEEEauthorblockN{P. N. Karthik and}
	\and
	\IEEEauthorblockN{Rajesh Sundaresan}
    \thanks{P. N. Karthik is with the Department of Electrical Communication Engineering, Indian Institute of Science, Bangalore, Karnataka 560012, India. Email: \{karti1992@ece.iisc.ernet.in\}. Rajesh Sundaresan is with the Department of Electrical Communication Engineering and the Robert Bosch Centre For Cyber-Physical Systems, Indian Institute of Science, Bangalore, Karnataka 560012, India. Email: \{rajeshs@ece.iisc.ernet.in\}}}
	\maketitle
	\providecommand{\keywords}[1]{\textbf{\textit{Index terms---}} #1}
	
	\begin{abstract}
	The aim of this work is to establish that two recently published projection theorems, one dealing with a parametric generalization of relative entropy and another dealing with R\'{e}nyi divergence, are equivalent under a correspondence on the space of probability measures. Further, we demonstrate that the associated ``Pythagorean'' theorems are equivalent under this correspondence. Finally, we apply Eguchi's method of obtaining Riemannian metrics from general divergence functions to show that the geometry arising from the above divergences are equivalent under the aforementioned correspondence.
    \end{abstract}

\section{Introduction}\label{sec:introduction}
Projection theorems fit into the following paradigm. Consider a space $ \mathbb{H} $ with a notion of a divergence $\mathscr{I}(P,Q)$ between any two points $P,Q\in \mathbb{H}$ that satisfies 
\begin{equation}
\mathscr{I}(P,Q)\geq 0,~\text{with equality if and only if }P=Q.\label{statement:defn_of_div_in_Hil_space}
\end{equation}
The projection of a point $Q$ onto a set $\mathbb{E}\subset \mathbb{H}$ is a member $P_{*}\in \mathbb{H}$ that satisfies
\begin{equation}
\mathscr{I}(P_{*},Q)=\inf\limits_{P\in \mathbb{E}}\mathscr{I}(P,Q),\label{eq:defn_of_proj_in_Hil_space}
\end{equation}
and may be viewed as the best approximant of $Q$ from the set $\mathbb{E}$. Projection theorems provide sufficient conditions on $\mathbb{E}$ for the existence and uniqueness of projections. 

(a) If $\mathbb{H}$ is a Hilbert space, $\mathscr{I}$ is the usual notion of distance $\left\langle P-Q,P-Q\right\rangle^{\frac{1}{2}}$ where $\left\langle\cdot,\cdot\right\rangle$ denotes the inner product, and if $\mathbb{E}$ is closed and convex, then a projection exists and is unique (see, for e.g., \cite[Ch. $11$, Th. $14$]{Bhatia2009}).\label{item:prob_in_bhatia}

(b) If $\mathbb{H}$ is the space of probability measures on an abstract measure space, $\mathscr{I}$ is the relative entropy, and $\mathbb{E}$ is convex and closed with respect to the total variation metric, then a projection exists and is unique (\cite[Th. $2.1$]{Csisz$cutea$r1975}).\label{item:prob_in_csiszar}

There are extensions in the latter context. 

(c) In \cite{Kumar2015}, $\mathbb{H}$ is the space of probability measures absolutely continuous with respect to some $\sigma$-finite measure $\mu$ and $\mathscr{I}$ is a parametric generalization of relative entropy, termed as relative $\alpha$-entropy, and denoted $\mathscr{I}_{\alpha}$ for $\alpha>0$, $\alpha\neq 1$; see Def. \ref{defn:rel_alp_ent} later. If $\mathbb{E}$ is convex and its corresponding set of $\mu$-densities is $L^{\alpha}(\mu)$-closed, then a projection exists and is unique (\cite[Th. $8$]{Kumar2015}).\label{item:prob_in_kumar_sundaresan}

(d) In \cite{Kumar2016}, $\mathbb{H}$ is as in (c) and $\mathscr{I}$ is the R\'{e}nyi divergence (see Def. \ref{defn:renyi_div} later) of order $\alpha$, denoted $D_{\alpha}$ and defined for $\alpha>0$, $\alpha\neq 1$. If $\mathbb{E}$ is $\alpha$-convex (see Def. \ref{defn:alpha_convex_set} later) and its corresponding set of $\mu$-densities is $L^{1}(\mu)$-closed, then a projection exists and is unique (\cite[Th. $1$]{Kumar2016}).\label{item:prob_in_kumar_sason}

From \cite[Lemma $2.(c)$]{Kumar2015}, we know that the relative $\alpha$-entropy between two probability measures is equal to the R\'{e}nyi divergence of order $1/\alpha$ between the corresponding $\alpha$-scaled measures (see Def. \ref{defn:escort_measure} later). This suggests that the hypotheses in (c) for the existence and uniqueness of projections for probability measures may be equivalent to those in (d) for the corresponding $\alpha$-scaled measures, with $\alpha$ in (d) replaced by $1/\alpha$. In this paper, we explore this connection, and prove the equivalence between the hypotheses in items (c) and (d) above. 

When $\mathbb{H}$ is the space of probability measures, relative $\alpha$-entropy satisfies a ``Pythagorean property'' that uniquely characterizes the projection \cite{Kumar2015}, \cite{Sundaresan2002}, \cite{Sundaresan2007}. Recently, van Erven and Harrem\"{o}es \cite{Erven2014} and \cite{Kumar2016} showed that an analogous property holds for R\'{e}nyi divergence. The authors of \cite{Erven2014} hinted at a plausible relation between their result with those of \cite{Sundaresan2002} and \cite{Sundaresan2007}. We argue in this paper that this is indeed the case, and show the equivalence between the Pythagorean theorems appearing in \cite{Kumar2015} and \cite{Erven2014}. 

For a probability measure $Q$ on a finite alphabet, \cite{Kumar2015} showed that when $\mathbb{E}$ is a linear family (see Def. \ref{defn:linear_family} later), the projection is a member of the $\alpha$-power-law family generated by $Q$ (see Def. \ref{defn:alpha_pow_law_family} later). Likewise, \cite{Kumar2016} showed that when $\mathbb{E}$ is an $\alpha$-linear family (see Def. \ref{defn:alp_lin_fam} later), the projection is a member of the $\alpha$-exponential family generated by $Q$ (see Def. \ref{defn:alpha_exp_family} later). We prove that (i) $\mathbb{E}$ is linear iff $\mathbb{E}^{(\alpha)}$, the set of $\alpha$-scaled measures associated with $\mathbb{E}$, is $(1/\alpha)$-linear, and (ii) the $\alpha$-power-law family generated by $Q$ is equivalent to the $(1/\alpha)$-exponential family generated by the $\alpha$-scaled measure of $Q$.

Towards the study of the geometric structure of statistical models under general divergences, Eguchi \cite{Eguchi1992} suggested a method of defining a {Riemannian metric} on {statistical manifolds} (see Def. \ref{defn:statistical_manifold} later) from a general divergence function. It is well known that Eguchi's method with relative entropy as the divergence results in a metric that is specified by the Fisher information matrix; see for example \cite[Sec. $2.2$ and Sec. $3.2$]{Amari2007}. We apply the same method to relative $\alpha$-entropy and R\'{e}nyi divergence, and show that under a suitable correspondence on the space of probability measures, the metrics specified in these cases are equivalent.

We set up the basic notation and definitions in Section \ref{sec:preliminaries}, present the main results in Section \ref{sec:main_results}, and some concluding remarks in Section \ref{sec:conclusions}. 

\section{Preliminaries}\label{sec:preliminaries}
Let $(\mathbb{X},\mathcal{X})$ be an abstract measure space, and let $\mu$ be any $\sigma$-finite measure on $(\mathbb{X},\mathcal{X})$. Let $P,Q$ be two probability measures absolutely continuous with respect to $\mu$, denoted $P\ll \mu$, $Q\ll \mu$. Let $p=\frac{dP}{d\mu}$ and $q=\frac{dQ}{d\mu}$ denote the respective $\mu$-densities. When $\mathbb{X}$ is finite, we take $\mu$ to be the counting measure. Consider $\alpha>0$, $\alpha\neq 1$. Throughout this paper, we assume that $p$ and $q$ belong to the complete topological vector space $L^{\alpha}(\mu)$ defined by the metric
\begin{align}
d(f,g)=
\begin{cases}
\left(\int |f-g|^{\alpha}d\mu\right)^{1/\alpha},~\alpha>1,\\
\int |f-g|^{\alpha}d\mu,~\alpha<1.
\end{cases} \label{eq:metric_L_alp}
\end{align}
We shall use the notation $||h||=\left(\int h^{\alpha}d\mu\right)^{1/\alpha}$, even though $||\cdot||$, as defined, is not a norm when $\alpha<1$. The dependence of $d(\cdot,\cdot)$ and $||\cdot||$ on $\alpha$ and $\mu$ is suppressed for convenience. 

\begin{defn}[$\alpha$-scaled measure]\label{defn:escort_measure}
Given a probability measure $P\ll \mu$ with $\mu$-density $p$, its $\alpha$-scaled measure $P^{(\alpha)}$ is the probability measure whose $\mu$-density $p^{(\alpha)}$ is
\begin{equation}
p^{(\alpha)}:=\frac{p^{\alpha}}{\int p^{\alpha}d\mu}=\left(\frac{p}{||p||}\right)^{\alpha}. \label{eq:escort_density}
\end{equation}	
\end{defn}

\begin{defn}[Relative $\alpha$-entropy]\label{defn:rel_alp_ent}
The relative $\alpha$-entropy of $P$ with respect to $Q$ is defined as
\begin{align}
\mathscr{I}_{\alpha}(P,Q)&:=\frac{\alpha}{1-\alpha}\log\left(\int \frac{p}{||p||}\left(\frac{q}{||q||}\right)^{\alpha-1}d\mu\right)\label{eq:rel_alpha_ent}.
\end{align}	
\end{defn}

\begin{defn}[R\'{e}nyi divergence]\label{defn:renyi_div}
The R\'{e}nyi divergence of order $\alpha$ between $P$ and $Q$ is defined as
\begin{equation}
D_{\alpha}(P||Q):=\frac{1}{\alpha-1}\log\left(\int p^{\alpha}q^{1-\alpha}d\mu\right)\label{eq:renyi_div_defn}.
\end{equation}	
\end{defn} 
The relation between relative $\alpha$-entropy and R\'{e}nyi divergence is known to be \cite[Lemma $2.(c)$]{Kumar2015} 
\begin{equation}
\mathscr{I}_{\alpha}(P,Q)=D_{1/\alpha}(P^{(\alpha)}||Q^{(\alpha)}),\label{eq:rel_betw_rel_alp_ent_and_Ren_div}
\end{equation}
where $P^{(\alpha)}$ and $Q^{(\alpha)}$ are the $\alpha$-scaled measures of $P$ and $Q$ respectively. For a subset $\mathbb{E}$ of probability measures absolutely continuous with respect to $\mu$, we denote the corresponding set of $\mu$-densities by $\mathcal{E}$, i.e.,
\begin{equation}
\mathcal{E}:=\left\lbrace p=\frac{dP}{d\mu}:P\in \mathbb{E} \right\rbrace. \label{eq:mu_densities}
\end{equation}
Also, we write $\mathbb{E}^{(\alpha)}$ for the set of $\alpha$-scaled measures associated with the probability measures in $\mathbb{E}$, and $\mathcal{E}^{(\alpha)}$ for its corresponding set of $\mu$-densities, i.e.,
\begin{equation}
\mathcal{E}^{(\alpha)}:=\left\lbrace p^{(\alpha)}=\frac{p^{\alpha}}{\int p^{\alpha}d\mu}:p\in \mathcal{E} \right\rbrace. \label{eq:escort_densities}
\end{equation}
It thus follows that the probability measures in $\mathbb{E}^{(\alpha)}$ are absolutely continuous with respect to $\mu$ whenever those in $\mathbb{E}$ are. Further, the sets $\mathbb{E}$ and $\mathbb{E}^{(\alpha)}$ are in one-one correspondence, i.e., for each $P\in \mathbb{E}$ such that $\frac{dP}{d\mu}=p$, there exists a unique $P^{(\alpha)}\in \mathbb{E}^{(\alpha)}$ such that its $\mu$-density $p^{(\alpha)}$ satisfies \eqref{eq:escort_density}. Conversely, for each $P^{(\alpha)}\in \mathbb{E}^{(\alpha)}$ such that $\frac{dP^{(\alpha)}}{d\mu}=p^{(\alpha)}$, there exists a unique $P\in \mathbb{E}$ (upto equivalence with respect to $\mu$) such that its $\mu$-density $p$ satisfies \eqref{eq:escort_density}.

\begin{defn}[The $p\longleftrightarrow p^{(\alpha)}$ correspondence]\label{defn:p_p'_corr}
Given a probability measure $P\ll \mu$ with $\mu$-density $p$, a function $p^{(\alpha)}$ is said to be in correspondence with $p$, denoted as $p\longleftrightarrow p^{(\alpha)}$, if it satisfies \eqref{eq:escort_density}.
\end{defn} 
On account of this definition, we have a one-one correspondence between $\mathcal{E}$ and $\mathcal{E}^{(\alpha)}$ whenever $\mathbb{E}$ and $\mathbb{E}^{(\alpha)}$ are in one-one correspondence.

\begin{defn}[$(\alpha,\lambda)-\text{mixture}$]\label{defn:alpha_lambda_mix}
Given two probability measures $P_{0},P_{1}\ll \mu$ and $\lambda\in (0,1)$, the $(\alpha,\lambda)$-mixture of $ P_{0} $ and $ P_{1} $ is the probability measure $P_{\alpha,\lambda}$ whose $\mu$-density $p_{\alpha,\lambda}$ is 

\begin{equation}
p_{\alpha,\lambda}:=\frac{\left(\lambda (p_{1})^{\alpha}+(1-\lambda)(p_{0})^{\alpha}\right)^{1/\alpha}}{\int \left(\lambda (p_{1})^{\alpha}+(1-\lambda)(p_{0})^{\alpha}\right)^{1/\alpha} d\mu}. \label{eq:alph_lam_mix}
\end{equation}
\end{defn}  
Note that $p_{\alpha,\lambda}$ is well-defined since 
\begin{equation}
Z:=\int \left(\lambda (p_{1})^{\alpha}+(1-\lambda)(p_{0})^{\alpha}\right)^{1/\alpha} d\mu
\end{equation} 
is always strictly positive and finite. Indeed, for $\lambda\in (0,1)$,
\begin{equation}
0 \leq \left(\lambda (p_{1})^{\alpha}+(1-\lambda)(p_{0})^{\alpha}\right)^{1/\alpha} \leq \max\{p_{0},p_{1}\} \leq p_{0}+p_{1}, \label{eq:finiteness_of_Z}
\end{equation}
which implies that $0 \leq Z \leq 2$. The first inequality in \eqref{eq:finiteness_of_Z} holds with equality if and only if $p_{1}\equiv 0$ and $p_{0}\equiv 0$. Hence, for any non-trivial densities $ p_{0} $ and $ p_{1} $, and hence for probability densities, we have $Z>0$.

\begin{defn}[$\alpha$-convex set]\label{defn:alpha_convex_set}
A set $\mathbb{E}$ of probability measures is said to be $\alpha$-convex if for any $P_{0},P_{1}\in \mathbb{E}$ and $\lambda\in (0,1)$, the $(\alpha,\lambda)$-mixture of $P_{0}$ and $P_{1}$ belongs to $\mathbb{E}$.
\end{defn}

\begin{defn}[$\alpha$-exponential function]\label{defn:alpha_exp_fun}
The $\alpha$-exponential function $e_{\alpha}:\mathbb{R}\cup \{\infty\}\rightarrow \mathbb{R}_{+}\cup \{\infty\}$ is defined as
\begin{align}
e_{\alpha}(u)=
\begin{cases}
(\max{\{1+(1-\alpha)u,0\}})^{\frac{1}{1-\alpha}},~&\alpha\neq 1,\\
\exp(u),~&\alpha=1.
\end{cases}  \label{eq:alpha_exp_fun}
\end{align}
\end{defn}
\begin{defn}[$\alpha$-power-law family]\label{defn:alpha_pow_law_family}
Given a probability measure $Q$ (with full support when $\alpha>1$), $k\in \{1,2,\ldots\}$ and $\Theta=\{\theta=(\theta_{1},\ldots,\theta_{k}):\theta_{i}\in \mathbb{R}\}\subset \mathbb{R}^{k}$, the $\alpha$-power-law family generated by $Q$ and functions $f_{i}:\mathbb{X}\rightarrow \mathbb{R}$, $1\leq i\leq k$, is defined as the set of probability measures
\begin{equation}
\mathcal{Z}^{(\alpha)}=\left\lbrace P_{\theta}:\theta\in \Theta\right\rbrace,
\end{equation}
where
\begin{equation}
P_{\theta}(x)^{-1}={M}(\theta)~e_{\alpha}\left(\frac{(Q(x))^{\alpha-1}-1}{1-\alpha}+\sum\limits_{i=1}^{k}\theta_{i}f_{i}(x)\right)\label{eq:alp_pow_law}
\end{equation}
\end{defn}
for $x\in \mathbb{X}$ with $M(\theta)$ being the normalisation constant. Assuming that the argument of $e_{\alpha}$ is strictly positive, using \eqref{eq:alpha_exp_fun} in \eqref{eq:alp_pow_law} yields
\begin{equation}
P_{\theta}(x)^{-1}=M(\theta)\left((Q(x))^{\alpha-1}+(1-\alpha)\sum\limits_{i=1}^{k}\theta_{i}f_{i}(x)\right)^{\frac{1}{1-\alpha}}\label{eq:alp_pow_law_family}
\end{equation} 
for $ x\in \mathbb{X} $.

\begin{defn}[$\alpha$-exponential family]\label{defn:alpha_exp_family}
Given a probability measure $Q$, $k\in \{1,2,\ldots\}$ and $\Theta=\{\theta=(\theta_{1},\ldots,\theta_{k}):\theta_{i}\in \mathbb{R}\}\subset \mathbb{R}^{k}$, the $\alpha$-exponential family generated by $Q$ and functions $f_{i}:\mathbb{X}\rightarrow \mathbb{R}$, $1\leq i\leq k$, is defined as the set of probability measures
\begin{equation}
\mathcal{Y}^{(\alpha)}=\left\lbrace P_{\theta}:\theta\in \Theta\right\rbrace,
\end{equation}
where
\begin{equation}
P_{\theta}(x)=(N(\theta))^{-1}\left((Q(x))^{1-\alpha}+(1-\alpha)\sum\limits_{i=1}^{k}\theta_{i}f_{i}(x)\right)^{\frac{1}{1-\alpha}}\label{eq:alp_exp_family}
\end{equation}
\end{defn}
for $x\in \mathbb{X}$, with ${N}(\theta)$ being the normalisation factor. The forms \eqref{eq:alp_pow_law_family} and \eqref{eq:alp_exp_family} will be used in Sec. \ref{subsec:alp_pow_law_exp}. 

\begin{defn}[Linear family]\label{defn:linear_family}
For any given functions $f_{1},\ldots,f_{k}$ on a finite alphabet $\mathbb{X}$, the family probability measures defined by
\begin{equation}
\mathbb{L}:=\left\lbrace P:\sum\limits_{x\in \mathbb{X}} f_{i}(x)P(x)=0,~1\leq i\leq k \right\rbrace, \label{eq:lin_family}
\end{equation}	
if nonempty, is called a linear family.
\end{defn}

\begin{defn}[$\alpha$-linear family]\label{defn:alp_lin_fam}
For any given functions $f_{1},\ldots,f_{k}$ on a finite alphabet $\mathbb{X}$, the family probability measures defined by
\begin{equation}
\mathbb{L}_{(\alpha)}:=\left\lbrace P:\sum\limits_{x\in \mathbb{X}} f_{i}(x)(P(x))^{\alpha}=0,~1\leq i\leq k \right\rbrace, \label{eq:alp_lin_family}
\end{equation}	
if nonempty, is called an $\alpha$-linear family. 
\end{defn}

Clearly, if $P$ is a member of $\mathbb{L}$, then $P^{(\alpha)}$ is a member of $\mathbb{L}_{(1/\alpha)}$. The converse likewise holds.

\begin{defn}[Statistical manifold]\label{defn:statistical_manifold}
A statistical manifold $S$ is a parametric family of probability distributions on $\mathbb{X}$ (with full support) with a continuously varying parameter space. It is usually represented as
\begin{equation}
S=\{p_{\phi}:\phi=(\phi_{1},\ldots,\phi_{n})\in\Phi\subset\mathbb{R}^{n}\}.\label{eq:statistical_manifolds}
\end{equation}
Here $\Phi$ is the parameter space. In writing \eqref{eq:statistical_manifolds}, we note that given any $\phi\in \Phi$, there exists a unique $p_{\phi}\in S$, and vice-versa. 
\end{defn}

The mapping $p\mapsto (\phi_{1}(p),\ldots,\phi_{n}(p))$ is called a \emph{coordinate system} for $S$. The \emph{tangent space} at a point $p$ on a manifold $S$, denoted as $T_{p}(S)$, is a linear space that corresponds to the linearization of the manifold around $p$; the elements of the tangent space are called \emph{tangent vectors}. For a coordinate system $\phi$, we denote the basis vectors of a tangent space $T_{p}(S)$ by $(\partial_{i})_{p}=(\partial/\partial\phi_{i})_{p}$, $i=1,\ldots,n$. A \emph{(Riemannian) metric} at a point $p\in S$ is an inner product defined between any two tangent vectors at that point. Although it is convenient to identify a metric with a point on the manifold, it is conventional to identify it with the coordinate $\phi(p)=(\phi_{1}(p),\ldots,\phi_{n}(p))$ of $p$. A metric is completely characterized by a matrix whose entries are the inner products between the basis tangent vectors, i.e., it is characterized by the matrix
\begin{equation}
G(\phi)=[g_{i,j}(\phi)]_{i,j=1,\ldots,n},\label{eq:matrx_specify_metric}
\end{equation}
where $g_{i,j}=\left\langle\partial_{i},\partial_{j}\right\rangle$.

Let $S$ be a manifold with a coordinate system $\phi=(\phi_{1},\ldots,\phi_{n})$, and let $\mathscr{I}$ be a divergence function on $S\times S$. We shall use the notation $\mathscr{I}(p,q)$ to denote the divergence $\mathscr{I}(P,Q)$ between the probability measures $P$ and $Q$. Eguchi \cite{Eguchi1992} showed that there is a metric
\begin{equation}
G^{(\mathscr{I})}(\phi)=[g^{(\mathscr{I})}_{i,j}(\phi)]_{i,j=1,\ldots,n}
\end{equation}
with   
\begin{equation}
g^{(\mathscr{I})}_{i,j}(\phi)=-\frac{\partial}{\partial\phi_{i}}\frac{\partial}{\partial\phi^{'}_{j}}\mathscr{I}(p_{\phi},p_{\phi^{'}})\biggr\rvert_{\phi^{'}=\phi},\label{eq:metric_for_gen_div}
\end{equation}
where $\phi=(\phi_{1},\ldots,\phi_{n})$ and $\phi^{'}=(\phi^{'}_{1},\ldots,\phi^{'}_{n})$. In Sec. \ref{subsec:riemannian_metrics}, we evaluate \eqref{eq:metric_for_gen_div} for the individual cases when $\mathscr{I}$ is either relative $\alpha$-entropy or R\'{e}nyi divergence, and thereafter demonstrate an equivalence between the two metrics.

\textit{Note:} For the material presented in Sec. \ref{subsec:alp_pow_law_exp} and Sec. \ref{subsec:riemannian_metrics}, we assume that $\mathbb{X}$ is a finite set.

\section{Main Results}\label{sec:main_results}
We begin this section with two important propositions that will be used to establish the results later.
\begin{prop}\label{prop:equiv_convex_alpha_convex}
Fix $\alpha>0$, $\alpha\neq 1$. A set $\mathbb{E}$ of probability measures absolutely continuous with respect to $\mu$ is convex if and only if the corresponding set of $\alpha$-scaled measures $\mathbb{E}^{(\alpha)}$ is $(1/\alpha)$-convex. 
\end{prop} 

\begin{IEEEproof}
See Appendix \ref{appndx:proof_of_Prop_1}.
\end{IEEEproof} 

\begin{prop}\label{prop:equiv_L1_Lalpha_closed}
Fix $\alpha>0$, $\alpha\neq 1$. Let $\mathbb{E}$ be a set of probability measures and let $\mathbb{E}^{(\alpha)}$ be the corresponding set of $\alpha$-scaled measures. Let $\mathcal{E}$ and $\mathcal{E}^{(\alpha)}$ be the set of $\mu$-densities associated with the probability measures in $\mathbb{E}$ and $\mathbb{E}^{(\alpha)}$ respectively. Then, $\mathcal{E}$ is closed in $L^{\alpha}(\mu)$ if and only if $\mathcal{E}^{(\alpha)}$ is closed in $L^{1}(\mu)$.
\end{prop}

\begin{IEEEproof}
See Appendix \ref{appndx:proof_of_prop_2}.
\end{IEEEproof}

\subsection{Equivalence of the Projection Problems}\label{sec:proj_probs}
We now consider the following two projection problems appearing in the works of \cite{Kumar2015} and \cite{Kumar2016} respectively:
\begin{enumerate}[(A)]
\item Fix $\alpha>0$, $\alpha\neq 1$. Let $Q$ be any probability measure, $Q\ll \mu$, and $\mathbb{E}$ be a set of probability measures whose set of $\mu$-densities is $\mathcal{E}$. Solve
\begin{equation}
\inf\limits_{P\in \mathbb{E}}\mathscr{I}_{\alpha}(P,Q).\label{eq:prob_in_kumar_sundaresan}
\end{equation}

\item Fix $\alpha>0$, $\alpha\neq 1$. Let $Q$ be any probability measure, $Q\ll \mu$, and $\mathbb{E}_{1}$ be a set of probability measures whose set of $\mu$-densities is $\mathcal{E}_{1}$. Solve
\begin{equation}
\inf\limits_{P\in \mathbb{E}_{1}}D_{\alpha}(P||Q).\label{eq:prob_in_kumar_sason}
\end{equation}
\end{enumerate}
Recall from Sec. \ref{sec:introduction}.(c) that a sufficient condition proposed in \cite{Kumar2015} for the existence and uniqueness of solution to \eqref{eq:prob_in_kumar_sundaresan} is that
\begin{equation}
\mathbb{E}~\text{is convex and }\mathcal{E}~\text{is closed in }L^{\alpha}(\mu),\label{eq:hypotheses_kumar_sundaresan}
\end{equation}
and from Sec. \ref{sec:introduction}.(d) that a sufficient condition proposed in \cite{Kumar2016} for the existence and uniqueness of solution to \eqref{eq:prob_in_kumar_sason} is that
\begin{equation}
\mathbb{E}_{1}~\text{is }\alpha\text{-convex and }\mathcal{E}_{1}~\text{is closed in }L^{1}(\mu).\label{eq:hypotheses_kumar_sason}
\end{equation}
We now demonstrate that, under the $p\longleftrightarrow p^{(\alpha)}$ correspondence, the problems \eqref{eq:prob_in_kumar_sundaresan} and \eqref{eq:prob_in_kumar_sason} are equivalent.

\begin{thrm}\label{thm:equiv_probs_sason_sundaresan}
The minimization problem \eqref{eq:prob_in_kumar_sundaresan} for a given $\alpha>0$, $\alpha\neq 1$, is equivalent to \eqref{eq:prob_in_kumar_sason} with $\alpha$ replaced by $1/\alpha$ and $\mathbb{E}_{1}$ replaced by $\mathbb{E}^{(\alpha)}$, the set of $\alpha$-scaled measures corresponding to $\mathbb{E}$. Moreover, the hypotheses in \eqref{eq:hypotheses_kumar_sundaresan} and \eqref{eq:hypotheses_kumar_sason} are identical under the $p\longleftrightarrow p^{(\alpha)}$ correspondence.
\end{thrm}

\begin{IEEEproof}
The problem in \eqref{eq:prob_in_kumar_sundaresan} is
\begin{equation}
\inf\limits_{P\in \mathbb{E}}\mathscr{I}_{\alpha}(P,Q).\label{eq:prob_in_kumar_sundaresan_repeat}
\end{equation}
Since \eqref{eq:rel_betw_rel_alp_ent_and_Ren_div} holds, under the $p\longleftrightarrow p^{(\alpha)}$ correspondence, the problem is equivalent to 
\begin{equation}
\inf\limits_{P^{(\alpha)}\in \mathbb{E}^{(\alpha)}}{D}_{1/\alpha}(P^{(\alpha)}||Q^{(\alpha)}),\label{eq:prob_in_kumar_sason_repeat}
\end{equation}
which is \eqref{eq:prob_in_kumar_sason}, with $\mathbb{E}_{1}$ replaced by $\mathbb{E}^{(\alpha)}$ and $\alpha$ replaced by $1/\alpha$. Further, by Props. \ref{prop:equiv_convex_alpha_convex} and \ref{prop:equiv_L1_Lalpha_closed}, the hypotheses in \eqref{eq:hypotheses_kumar_sundaresan} and \eqref{eq:hypotheses_kumar_sason} are equivalent, with $\mathcal{E}_{1}$ replaced by $\mathcal{E}^{(\alpha)}$.
\end{IEEEproof}

\subsection{Equivalence of the Pythagorean Theorems}\label{subsec:pythagorean_property}
We now argue the equivalence between the theorems on the ``Pythagorean property'' of relative $\alpha$-entropy and R\'{e}nyi divergence.	The result \cite[Th. $10.(a)$]{Kumar2015} establishes that if $\mathbb{E}$ is convex, then the projection $P_{*}$ of $Q$ onto $\mathbb{E}$, if it exists, satisfies
\begin{equation}
\mathscr{I}_{\alpha}(P,Q)\geq \mathscr{I}_{\alpha}(P,P_{*}) + \mathscr{I}_{\alpha}(P_{*},Q)~\text{for all } P\in \mathbb{E}.\label{eq:pyt_ineq_rel_alp_ent}
\end{equation}
By virtue of $(1/\alpha)$-convexity of $ \mathbb{E}^{(\alpha)} $ (Proposition \ref{prop:equiv_convex_alpha_convex}) and \eqref{eq:rel_betw_rel_alp_ent_and_Ren_div}, $P_{*}^{(\alpha)}$ is the $D_{1/\alpha}$-projection of $Q^{(\alpha)}$ onto $\mathbb{E}^{(\alpha)}$ and this projection satisfies
\begin{equation}
{D}_{1/\alpha}(P^{(\alpha)},Q^{(\alpha)})\geq {D}_{1/\alpha}(P^{(\alpha)},P^{(\alpha)}_{*}) + {D}_{1/\alpha}(P^{(\alpha)}_{*},Q^{(\alpha)})\label{eq:pyt_ineq_ren_div}
\end{equation}
$\text{for all } P^{(\alpha)}\in \mathbb{E}^{(\alpha)}$. This recovers \cite[Th. $14$]{Erven2014}, as also \cite[Prop. $1$]{Kumar2016}, with $1/\alpha$ replacing $\alpha$.

\subsection{Equivalence of $\alpha$-Power-Law and $\alpha$-Exponential Families}\label{subsec:alp_pow_law_exp}
We now demonstrate that, under the $p\longleftrightarrow p^{(\alpha)}$ correspondence, the $\alpha$-power-law family generated by a probability measure $Q$ is equivalent to the ${\alpha}$-exponential family generated by the $\alpha$-scaled measure of $Q$. 
\begin{thrm}\label{thm:alp_pow_law_alp_exp}
Let $\mathbb{X}$ be a finite alphabet. Fix $\alpha>0$, $\alpha\neq 1$, $k\in \{1,2,\ldots\}$, and $\Theta=\{\theta=(\theta_{1},\ldots,\theta_{k}):\theta_{i}\in \mathbb{R}\}\subset \mathbb{R}^{k}$. Let $f_{i}:\mathbb{X}\rightarrow \mathbb{R}$, $1\leq i\leq k$, be specified. Given a probability measure $Q$, for every member of the $\alpha$-power-law family generated by $Q$, $f_{1},\ldots,f_{k}$ and $\Theta$, its $\alpha$-scaled measure is a member of the $(1/\alpha)$-exponential family generated by $Q^{(\alpha)}$, $f_{1},\ldots,f_{k}$ and $\Theta'$, where $\Theta'$ is a scalar modification of $\Theta$ that depends on $Q$.  
\end{thrm}
\begin{IEEEproof}
See Appendix \ref{appndx:proof_of_equiv_alp_pow_law_alp_exp}.
\end{IEEEproof}

\subsection{Equivalence of the Riemannian Metrics for Relative $\alpha$-Entropy and R\'{e}nyi Divergence}\label{subsec:riemannian_metrics}
It is well known from \cite[Sec. $2.2$ and Sec. $3.2$]{Amari2007} that when $\mathscr{I}(p,q)=I(p||q)$, the relative entropy between $p$ and $q$, \eqref{eq:metric_for_gen_div} can be written as
\begin{align}
g^{({I})}_{i,j}(\phi)&=E_{p_{\phi}}[\partial_{i}\log p_{\phi},\partial_{j}\log p_{\phi}],\label{eq:matrix_for_rel_ent_metric}
\end{align} 
where $E_{p_{\phi}}$ denotes the expectation with respect to $p_{\phi}$. The quantity in \eqref{eq:matrix_for_rel_ent_metric} is the $(i,j)$ entry of the Fisher information matrix. Thus, with relative entropy as the divergence function, the Riemannian metric is the one specified by the Fisher information matrix.

	On similar lines, when $\mathscr{I}(p,q)=D_{\alpha}(p||q)$, the R\'{e}nyi divergence of order $\alpha$ between $p$ and $q$, where $\alpha>0$, $\alpha\neq 1$,
	using \eqref{eq:renyi_div_defn} in \eqref{eq:metric_for_gen_div} for the finite alphabet setting results in the following set of equations:
	\begin{align}
	g^{(D_{\alpha})}_{i,j}(\phi)&=-\frac{\partial}{\partial\phi_{i}}\frac{\partial}{\partial\phi^{'}_{j}}D_{\alpha}(p_{\phi}||p_{\phi^{'}})\biggr\rvert_{\phi^{'}=\phi}\nonumber\\
	                 &=-\frac{1}{\alpha-1}\cdot\frac{\partial}{\partial\phi_{i}}\frac{\partial}{\partial\phi^{'}_{j}}\log\left(\sum\limits_{x\in \mathbb{X}}p_{\phi}(x)^{\alpha}p_{\phi^{'}}(x)^{1-\alpha}\right)\biggr\rvert_{\phi^{'}=\phi}\\
	                 &=\frac{\partial}{\partial\phi_{i}}\left(\frac{\sum\limits_{x\in \mathbb{X}}p_{\phi}(x)^{\alpha}p_{\phi^{'}}(x)^{-\alpha}~\partial_{j}^{'}p_{\phi^{'}}(x)}{\sum\limits_{x\in \mathbb{X}}p_{\phi}(x)^{\alpha}p_{\phi^{'}}(x)^{1-\alpha}}\right)\biggr\rvert_{\phi^{'}=\phi}\\
	                 &=\alpha\biggr(\sum\limits_{x\in \mathbb{X}}\partial_{i}p_{\phi}(x)\cdot \partial_{j}\log p_{\phi}(x)
	                  -E_{\phi}[\partial_{i}\log p_{\phi}]\cdot E_{\phi}[\partial_{j}\log p_{\phi}]\biggr)\\
	                 &=\alpha\biggr(\sum\limits_{x\in \mathbb{X}}p_{\phi}(x)\cdot\partial_{i} \log p_{\phi}(x)\cdot  \partial_{j}\log p_{\phi}(x)\biggr)\label{eq:expec_score_fun_zero_once_again}\\
	                 &=\alpha\cdot E_{p_{\phi}}[\partial_{i} \log p_{\phi}\cdot \partial_{j}\log p_{\phi}]\\
	                 &=\alpha\cdot g^{(I)}_{i,j}(\phi)\label{eq:matrix_renyi_div_metric}\\
	                 &=g_{i,j}^{(\alpha I)}(\phi)\label{eq:matrix_alp_rel_alp_ent_metric}, 
	\end{align}
	where \eqref{eq:expec_score_fun_zero_once_again} follows by using the fact that the expectation of the score function is zero, \ref{eq:matrix_renyi_div_metric} follows from \eqref{eq:matrix_for_rel_ent_metric}, and \eqref{eq:matrix_alp_rel_alp_ent_metric} follows by recognizing that \eqref{eq:matrix_renyi_div_metric} can be obtained from \eqref{eq:metric_for_gen_div} by plugging $\mathscr{I}(p,q)=\alpha I(p,q)$.
	Our next result explores the consequences of the $p\longleftrightarrow p^{(\alpha)}$ correspondence.
	
	\begin{thrm}\label{thm:equiv_metrics_rel_alp_ent_ren_div}
		Consider a finite alphabet $\mathbb{X}$ and fix $\alpha>0$, $\alpha\neq 1$. Let $S$ be a statistical manifold equipped with a coordinate system $\phi=(\phi_{1},\ldots,\phi_{n})$, and let $S^{(\alpha)}$ denote the statistical manifold of the corresponding $\alpha$-scaled measures. Then, for every $p\in S$, the Riemannian metric specified by relative $\alpha$-entropy on $T_{p}(S)$ is equivalent to that specified by R\'{e}nyi divergence of order $1/\alpha$ on $T_{p^{(\alpha)}}(S^{(\alpha)})$.
	\end{thrm}
	
	\begin{IEEEproof}
		The proof is immediate from \eqref{eq:rel_betw_rel_alp_ent_and_Ren_div}.
	\end{IEEEproof}
	
	By virtue of \eqref{eq:matrix_alp_rel_alp_ent_metric} and the above theorem, we can also conclude that for every $p\in S$, the Riemannian metric specified by relative $\alpha$-entropy on $T_{p}(S)$ is equivalent to that specified by $\alpha^{-1}I$ on $T_{p^{(\alpha)}}(S^{(\alpha)})$, where $I$ is the relative entropy.

	\section{Conclusions}\label{sec:conclusions}
	Owing to the correspondence $p\longleftrightarrow p^{(\alpha)}$, several independently established parallel results can now be viewed through a single lens.
	
	\section{Acknowledgment}
	This work was supported by the Robert Bosch Centre for Cyber-Physical Systems at the Indian Institute of Science.
	
	\appendix
	
	\subsection{Proof of Proposition \ref{prop:equiv_convex_alpha_convex}}\label{appndx:proof_of_Prop_1}
	We begin with the ``only if'' part. Suppose that $\mathbb{E}$ is a convex set of probability measures. Let $P_{0}^{(\alpha)},P_{1}^{(\alpha)}\in \mathbb{E}^{(\alpha)}$ and $\lambda\in (0,1)$ be arbitrary. Let $P^{(\alpha)}_{\frac{1}{\alpha},\lambda}$ denote the $\left(\frac{1}{\alpha},\lambda\right)$-mixture of $P_{0}^{(\alpha)}$ and $P_{1}^{(\alpha)}$. Then, we need to show that ${dP^{(\alpha)}_{\frac{1}{\alpha},\lambda}}/{d\mu}=p^{(\alpha)}_{\frac{1}{\alpha},\lambda}\in \mathcal{E}^{(\alpha)}$. Using \eqref{eq:alph_lam_mix}, we have	
	\begin{align}
	p^{(\alpha)}_{\frac{1}{\alpha},\lambda}&\propto
	{\left(\lambda (p_{1}^{(\alpha)})^{1/\alpha}+(1-\lambda)(p_{0}^{(\alpha)})^{1/\alpha}\right)^{\alpha}} \\
	&\propto{\left(\frac{\lambda}{||p_{1}||} {p_{1}}+\frac{(1-\lambda)}{||p_{0}||}{p_{0}}\right)^{\alpha}}\label{eq:replace_'_by_defn}\\
	&\propto{\left(\lambda' {p_{1}}+(1-\lambda'){p_{0}}\right)^{\alpha}},\label{eq:normalisation}
	\end{align}
	where \eqref{eq:replace_'_by_defn} follows from the application of \eqref{eq:escort_density} to $p_{1}^{(\alpha)}$ and $p_{0}^{(\alpha)}$ and \eqref{eq:normalisation} follows by setting
	\begin{equation}
	\lambda'=\frac{\frac{\lambda}{||p_{1}||}}{\frac{\lambda}{||p_{1}||}+\frac{1-\lambda}{||p_{0}||}}\label{eq:lambda'}
	\end{equation}
	and then absorbing the scaling in the normalisation constant. We now recognize that	
	\begin{equation}
	\lambda' p_{1}+ (1-\lambda')p_{0}=\frac{d\left(\lambda' P_{1}+ (1-\lambda')P_{0}\right)}{d\mu},
	\end{equation}
	and since $\mathbb{E}$ is convex by assumption, we have $\lambda' P_{1}+ (1-\lambda')P_{0}\in \mathbb{E}$, which implies, by \eqref{eq:mu_densities}, that $ \lambda' p_{1}+ (1-\lambda')p_{0}\in \mathcal{E} $. Using this and the fact that \eqref{eq:normalisation} implies $p^{(\alpha)}_{\frac{1}{\alpha},\lambda}\longleftrightarrow (\lambda' p_{1}+ (1-\lambda')p_{0})$, we conclude that $p^{(\alpha)}_{\frac{1}{\alpha},\lambda}\in \mathcal{E}^{(\alpha)}$, hence completing the proof of the ``only if'' part.
	
	We now proceed to prove the ``if part''. Suppose that $\mathbb{E}^{(\alpha)}$ is $(1/\alpha)$-convex. We need to show that for any $P_{0},P_{1}\in \mathbb{E}$ and $\lambda\in (0,1)$, we have $ \lambda P_{1} + (1-\lambda)P_{0}\in \mathbb{E} $. By definition, $P_{0}^{(\alpha)},P_{1}^{(\alpha)}\in \mathbb{E}^{(\alpha)}$. Let $p^{(\alpha)}_{0}=dP^{(\alpha)}_{0}/d\mu$, $p^{(\alpha)}_{1}=dP^{(\alpha)}_{1}/d\mu$. Set 
	\begin{equation}
	\lambda''=\frac{\frac{\lambda}{||p_{0}||}}{\frac{\lambda}{||p_{0}||}+\frac{1-\lambda}{||p_{1}||}}\label{eq:lambda''}.
	\end{equation} 
	Noting that $\lambda''\in (0,1)$ and $\mathbb{E}^{(\alpha)}$ is $(1/\alpha)$-convex, the $\left(\frac{1}{\alpha},\lambda''\right)$-mixture of $P_{0}^{(\alpha)}$ and $P_{1}^{(\alpha)}$ belongs to $\mathbb{E}^{(\alpha)}$. This implies that	
	\begin{align}
	p^{(\alpha)}_{\frac{1}{\alpha},\lambda''}&\propto{\left(\lambda'' (p_{1}^{(\alpha)})^{1/\alpha}+(1-\lambda'')(p_{0}^{(\alpha)})^{1/\alpha}\right)^{\alpha}}\nonumber\\
	&\propto{\left(\lambda'' \frac{p_{1}}{||p_{1}||}+(1-\lambda'')\frac{p_{0}}{||p_{0}||}\right)^{\alpha}}\nonumber\\
	&\propto{\left(\lambda {p_{1}}+(1-\lambda){p_{0}}\right)^{\alpha}}\label{eq:subst_for_lambda''}
	\end{align}
	belongs to $\mathcal{E}^{(\alpha)}$, where \eqref{eq:subst_for_lambda''} follows by plugging in \eqref{eq:lambda''} for $\lambda''$. Since \eqref{eq:subst_for_lambda''} implies $p^{(\alpha)}_{\frac{1}{\alpha},\lambda''}\longleftrightarrow (\lambda p_{1}+ (1-\lambda)p_{0})$, we conclude that $(\lambda p_{1}+(1-\lambda)p_{0})\in \mathcal{E}$, which implies, by \eqref{eq:mu_densities}, that $\lambda P_{1}+(1-\lambda)P_{0}\in \mathbb{E}$, hence completing the proof of the ``if'' part.
	
	\subsection{Proof of Proposition \ref{prop:equiv_L1_Lalpha_closed}}\label{appndx:proof_of_prop_2}
	The arguments we present here are already in \cite{Kumar2015}, but not in an isolated form. We bring them out here to establish the centrality of the correspondence. 
	
	We prove the forward and backward directions in order.
	\begin{enumerate}[(i)]
	\item $\Longrightarrow:$ Let $\mathcal{E}$ be closed in $L^{\alpha}(\mu)$. Let $p^{(\alpha)}$ be any limit point of $\mathcal{E}^{(\alpha)}$. If $p^{(\alpha)}\in \mathcal{E}^{(\alpha)}$, then there is nothing to prove. So, suppose that $p^{(\alpha)}\notin \mathcal{E}^{(\alpha)}$. Then, there exists a sequence $\{p_{n}^{(\alpha)}\}\subset \mathcal{E}^{(\alpha)}$ such that $p_{n}^{(\alpha)}\rightarrow p^{(\alpha)}$ in $L^{1}(\mu)$, i.e.,
	\begin{equation}
	\lim\limits_{n\rightarrow \infty} \int |p_{n}^{(\alpha)}-p^{(\alpha)}|d\mu = 0.\label{eq:L1_mu_convergence}
	\end{equation}
	It follows that $\int p_{n}^{(\alpha)} d\mu \rightarrow \int p^{(\alpha)} d\mu$, and since $\int p_{n}^{(\alpha)}d\mu=1$ for all $n$, we must have $\int p^{(\alpha)}d\mu=1$. 
	
	From the $L^{1}(\mu)$ convergence in \eqref{eq:L1_mu_convergence}, it follows that $p_{n}^{(\alpha)}\rightarrow p^{(\alpha)}$ in $[\mu]$-measure. We now demonstrate that the $\mu$-density proportional to $(p^{(\alpha)})^{1/\alpha}$ is in $\mathcal{E}$, thereby establishing that $p^{(\alpha)}\in \mathcal{E}^{(\alpha)}$ and hence the fact that $\mathcal{E}^{(\alpha)}$ is closed. 
	
	In view of the convergence in $[\mu]$-measure and the upper bound 
	\begin{equation}
	|(p_{n}^{(\alpha)})^{1/\alpha}-(p^{(\alpha)})^{1/\alpha}|^{\alpha}\leq 2^{\alpha}(p_{n}^{(\alpha)}+p^{(\alpha)}),
	\end{equation} 
	we can apply the generalized version of the dominated convergence theorem (see \cite{Folland1999}, Ch. $2$, Ex. $20$) to get 
	\begin{equation}
	\frac{p_{n}}{||p_{n}||}=(p_{n}^{(\alpha)})^{1/\alpha}\longrightarrow (p^{(\alpha)})^{1/\alpha}~\text{in }L^{\alpha}(\mu).
	\end{equation}
	We now claim that 
	\begin{equation}
	||p_{n}||~\text{is bounded.}\label{eq:norm_pn_bounded}
	\end{equation} 
	Suppose not; then working on a subsequence if needed, we have $||p_{n}||:=M_{n}\rightarrow \infty$. As $\int p_{n}d\mu=1$, given any $\epsilon>0$,
	\begin{align}
	\mu\left(p_{n}^{(\alpha)}>\epsilon\right)&=\mu\left(p_{n}>\epsilon^{1/\alpha}M_{n}\right)\nonumber\\
	&\leq \frac{1}{\epsilon^{1/\alpha}M_{n}}\longrightarrow 0~\text{as }n\rightarrow \infty,   
	\end{align}
	and hence $p_{n}^{(\alpha)}\rightarrow 0$ in $[\mu]$-measure, or $p^{(\alpha)}=0$ except on a set of $[\mu]$-measure 0 (i.e., $p^{(\alpha)}=0$ a.e.$[\mu]$). But this is a contradiction since $\int p^{(\alpha)}d\mu=1$. Hence, \eqref{eq:norm_pn_bounded} holds, and we can pick a subsequence of the sequence $||p_{n}||$ that converges to some $c>0$. Reindex
	and work on this subsequence to get $p_{n}\rightarrow c(p^{(\alpha)})^{1/\alpha}$ in $L^{\alpha}(\mu)$.
	The closedness of $\mathcal{E}$ implies that the limiting function $c(p^{(\alpha)})^{1/\alpha}=q$ for some $q\in \mathcal{E}$. Since we also have $\int p^{(\alpha)}d\mu=1$, it follows that $c=||q||$ and $p^{(\alpha)}=\left(q/||q||\right)^{\alpha}$. Thus, we have $p^{(\alpha)}\longleftrightarrow q$, which implies that $p^{(\alpha)}\in \mathcal{E}^{(\alpha)}$. This completes the proof of one direction.
	
	\item $ \Longleftarrow :$ Suppose that $\mathcal{E}^{(\alpha)}$ is closed in $L^{1}(\mu)$. Let $p$ be any arbitrary limit point of $\mathcal{E}$. Following the arguments as before, if $p\in \mathcal{E}$, then there is nothing to prove. So, suppose that $p\notin \mathcal{E}$. Then, there exists a sequence $\{p_{n}\}\subset \mathcal{E}$ such that $p_{n}\rightarrow p$ in $L^{\alpha}(\mu)$, i.e.,
	\begin{equation}
	\lim\limits_{n\rightarrow \infty} \int |p_{n}-p|^{\alpha}d\mu = 0.\label{eq:L_alpha_mu_convergence}
	\end{equation}
	This also implies that $||p_{n}||\rightarrow ||p||>0$, and since $|p_{n}^{\alpha}-p^{\alpha}|\leq p_{n}^{\alpha}+p^{\alpha}$, the generalized version of the dominated convergence theorem (\cite{Folland1999}, Ch. $2$, Ex. $20$) yields
	\begin{equation}
	p_{n}^{(\alpha)}=\left({p_{n}}/{||p_{n}||}\right)^{\alpha}\longrightarrow \left({p}/{||p||}\right)^{\alpha}~\text{in }L^{1}(\mu).
	\end{equation}
	The closedness of $\mathcal{E}^{(\alpha)}$ in $L^{1}(\mu)$ implies that the limiting function $\left({p}/{||p||}\right)^{\alpha}=p^{(\alpha)}$ for some $p^{(\alpha)}\in \mathcal{E}^{(\alpha)}$. This implies that $p\longleftrightarrow p^{(\alpha)}$, and thus the fact that $p\in \mathcal{E}$, thereby demonstrating that $\mathcal{E}$ is closed in $L^{\alpha}(\mu)$.
	\end{enumerate}
	
	\subsection{Proof of Theorem \ref{thm:alp_pow_law_alp_exp}}\label{appndx:proof_of_equiv_alp_pow_law_alp_exp}
	Suppose that $P_{\theta}$. $\theta\in \Theta$, is a member of the $\alpha$-power-law family generated by $Q$. According to \eqref{eq:alp_pow_law_family}, for any $x\in \mathbb{X}$, we have
	\begin{align}
	P_{\theta}(x)\propto {\left((Q(x))^{\alpha-1}+(1-\alpha)\sum\limits_{i=1}^{k}\theta_{i}f_{i}(x)\right)^{-\frac{1}{1-\alpha}}}.
	\end{align}
	From this, we get
	\begin{align}
	P^{(\alpha)}_{\theta}(x)&\propto{(P_{\theta}(x))^{\alpha}}\nonumber\\
	&\propto {\left((Q(x))^{\alpha-1}+(1-\alpha)\sum\limits_{i=1}^{k}\theta_{i}f_{i}(x)\right)^{-\frac{\alpha}{1-\alpha}}}\\
	&\propto{\left(\left(\frac{Q(x)}{||Q||}\right)^{{\alpha-1}}+(1-\alpha)\sum\limits_{i=1}^{k}\frac{\theta_{i}}{||Q||^{{\alpha-1}}}f_{i}(x)\right)^{-\frac{\alpha}{1-\alpha}}}\label{eq:alp_exp_fam_escort_measure_previous}\\
	&\propto{\left((Q^{(\alpha)}(x))^{1-\frac{1}{\alpha}}+\left(1-\frac{1}{\alpha}\right)\sum\limits_{i=1}^{k}\theta'_{i}f_{i}(x)\right)^{\frac{1}{1-\frac{1}{\alpha}}}},\label{eq:alp_exp_fam_escort_measure}
	\end{align}
	where \eqref{eq:alp_exp_fam_escort_measure_previous} follows by multiplying the scale factor $||Q||^{\alpha}$ and  \eqref{eq:alp_exp_fam_escort_measure} follows by setting $\theta'_{i}:=\frac{(-\alpha)\theta_{i}}{||Q||^{\alpha-1}}$, $1\leq i\leq k$. We recognize that \eqref{eq:alp_exp_fam_escort_measure} is of the form \eqref{eq:alp_exp_family}, with $\alpha$ replaced by $1/\alpha$, $Q(x)$ replaced by $Q^{(\alpha)}(x)$, and $\theta_{i}$ replaced by $\theta'_{i}$. This completes the proof.		 
	 
	\bibliographystyle{IEEEtran}
	\bibliography{IEEEabrv,arxiv_version}

\begin{thebibliography}{10}
\providecommand{\url}[1]{#1}
\csname url@samestyle\endcsname
\providecommand{\newblock}{\relax}
\providecommand{\bibinfo}[2]{#2}
\providecommand{\BIBentrySTDinterwordspacing}{\spaceskip=0pt\relax}
\providecommand{\BIBentryALTinterwordstretchfactor}{4}
\providecommand{\BIBentryALTinterwordspacing}{\spaceskip=\fontdimen2\font plus
\BIBentryALTinterwordstretchfactor\fontdimen3\font minus
  \fontdimen4\font\relax}
\providecommand{\BIBforeignlanguage}[2]{{%
\expandafter\ifx\csname l@#1\endcsname\relax
\typeout{** WARNING: IEEEtran.bst: No hyphenation pattern has been}%
\typeout{** loaded for the language `#1'. Using the pattern for}%
\typeout{** the default language instead.}%
\else
\language=\csname l@#1\endcsname
\fi
#2}}
\providecommand{\BIBdecl}{\relax}
\BIBdecl

\bibitem{Bhatia2009}
R.~Bhatia, \emph{Notes on functional analysis}.\hskip 1em plus 0.5em minus
  0.4em\relax Hindustan Book Agency, 2009.

\bibitem{Csisz$cutea$r1975}
I.~Csisz{$\acute{a}$}r, ``I-divergence geometry of probability distributions
  and minimization problems,'' \emph{Annals of Probability}, vol.~3, pp.
  146--158, 1975.

\bibitem{Kumar2015}
M.~A. Kumar and R.~Sundaresan, ``Minimization problems based on relative
  {$\alpha$}-entropy {$\text{I}$}: Forward projection,'' \emph{IEEE
  Transactions on Information Theory}, vol.~61, no.~9, pp. 5063--5080,
  September 2015.

\bibitem{Kumar2016}
M.~A. Kumar and I.~Sason, ``Projection theorems for the
  {$\text{R}$}{$\acute{e}$}nyi divergence on {$\alpha$}-convex sets,''
  \emph{IEEE Transactions on Information Theory}, vol.~62, no.~9, pp.
  4924--4935, September 2016.

\bibitem{Sundaresan2002}
R.~Sundaresan, ``A measure of discrimination and its geometric properties,'' in
  \emph{Information Theory, 2002. Proceedings. 2002 IEEE International
  Symposium on}.\hskip 1em plus 0.5em minus 0.4em\relax IEEE, 2002, p. 264.

\bibitem{Sundaresan2007}
------, ``Guessing under source uncertainty,'' \emph{IEEE Trans. on Information
  Theory}, vol.~53, no.~1, pp. 269--287, January 2007.

\bibitem{Erven2014}
T.~v. Erven and P.~Harremo{$\ddot{e}$}s, ``R{$\acute{e}$}nyi divergence and
  kullback-leibler divergence,'' \emph{IEEE Trans. on Information Theory},
  vol.~60, no.~7, pp. 3797--3820, July 2014.

\bibitem{Eguchi1992}
S.~Eguchi, ``Geometry of minimum contrast,'' \emph{Hiroshima Math. J}, vol.~22,
  no.~3, pp. 631--647, 1992.

\bibitem{Amari2007}
S.-i. Amari and H.~Nagaoka, \emph{Methods of information geometry}.\hskip 1em
  plus 0.5em minus 0.4em\relax American Mathematical Soc., 2007, vol. 191.

\bibitem{Folland1999}
G.~B. Folland, \emph{Real Analysis: Modern Techniques and Their Applications},
  2nd~ed.\hskip 1em plus 0.5em minus 0.4em\relax New York, NY, USA: Wiley,
  1999.

\end{thebibliography}
	\end{document}